# A General Formulation for Evaluating the Performance of Linear Power Flow Models

Zhentong Shao, *Student Member, IEEE*, Qiaozhu Zhai, *Member, IEEE*, and Xiaohong Guan, *Fellow, IEEE*

*Abstract*—Linear power flow (LPF) models are essential in power system analysis. Various LPF models are proposed, but some crucial questions are still remained: what is the performance bound (e.g., the error bound) of LPF models, how to know a branch is applicable for LPF models or not, and what is the best LPF model. In this paper, these crucial questions are answered and a general formulation (GF) for evaluating the performance of LPF models is proposed. The GF actually figure out two core difficulties, the one is how to define the definition range of the LPF models, and the second is how to analytically obtain the best LPF model and evaluate the performance of a given LPF model. Besides, the key factors that affect the performance of LPF models are also analyzed through the proposed framework. The case studies compare the proposed LPF model with the DC power flow model, the physical-model-driven LPF model, and the data-driven LPF model, and the results show the effectiveness as well as the superiority of the proposed method.

## Nomenclature

### A. Indices and Sets

$n, i, j$   Index of system buses, $n, i, j \in \{1, 2, \cdots, N\}$.
$k$   Index of data samples/scenarios, $k \in \{1, 2, \cdots, K\}$.
$ij$   Index of the branch from bus $i$ to bus $j$.
$l$   Index of the branch with the ends of bus $i$ and bus $j$, it has $l \in \{ij, ji\}$.
AS   The set of all possible solutions of the AC-PF equations.
RS   The set of all possible solutions of the AC-PF equations that have a high probability in engineering practice.
HS   The set of all possible solutions of the AC-PF equations that have appeared in the history.

### B. Parameters

$\varphi_l$   Impedance angle of branch $l$.
$y_l / z_l$   Admittance/Impedance of branch $l$.
$g_l / b_l$   Conductance/Susceptance of branch $l$.
$r_l / x_l$   Resistance/Reactance of branch $l$.
$\overline{F}_l^p / \overline{F}_l^q$   Active/Reactive power limitation of branch $l$.
$\overline{V}_i / \underline{V}_i$   Upper/Lower bound of voltage magnitude at bus $i$.
$\underline{\theta}_{ij} / \overline{\theta}_{ij}$   Upper/Lower bound of the angle across branch $(i, j)$.

### C. Variables

$\alpha$   Constant term of the LPF model.
$\beta / \gamma$   Linear factor of voltage magnitude/angle in the LPF model.
$p_{ij} / q_{ij}$   Active/Reactive power flow from bus $i$ to bus $j$.
$p_l / q_l$   Active/Reactive power flow of branch $l$.
$v_i / \theta_i$   Voltage magnitude/angle at bus $i$.

### D. Functions

$\delta(\cdot)$   Function that represent error metrics.
$\chi(v)$   Function that represents the variable space of voltage magnitudes.
$\phi(\theta)$   Function that represents the variable space of voltage angles.

## I. Introduction

THE AC branch power flow (AC-PF) model accurately describe the physical laws of steady-state AC power grid, which is essential in power system analysis. However, the AC-PF model is mathematically nonlinear and nonconvex, which leads to non-convergence and computational burdens in optimization problems, such as the problems of optimal power flow [1], congestion analysis [2], unit commitment [3] and economic dispatch [4]. To address the obstacles of AC-PF model, various linear power flow (LPF) models are proposed to give a linear approximation of the AC-PF model [5, 6].

Although numerous types of LPF models are proposed, some crucial questions related to the LPF models are still remained: 1) What is the performance bound (e.g., the error bound) of LPF models? 2) How to know a branch is applicable for LPF models or not? 3) What is the best LPF model, and how to obtained it?

In this paper, we are trying to answer these questions. Before presenting the answers, the review of the existing LPF models is given. The up-to-date LPF models can be classified into three types, which are the DC power flow (DC-PF) model, the physical-model-driven LPF (P-LPF) model and the data-driven LPF (D-LPF) model.

The DC-PF model is a traditional but widely used LPF model. The classical DC-PF model originates from engineering in early days [7], and the term "DC" comes from the assumption of ignoring reactive power flows. In literature [7], a thoroughly review is given to classify various DC-PF models into a cold-start type and a hot-start type. The hot-start DC-PF models improve the approximation accuracy by adopting empirical base points [8], and when the base points are unavailable, the cold-start DC-PF models are used [3, 9]. Beyond the review of [7], the literature [10] extends the DC-PF by considering fuzzy-variable-based power injections, the literature [11] analyzes the error bound of DC-PF through a convex relaxation approach; literatures [12, 13] focus on the influence of topology on the approximation accuracy of the DC-PF model, and gives a new

This work is supported in part by the National Natural Science Foundation (61773309, 61773308, 11991023, 11991020) of China. (*Corresponding author: Qiaozhu Zhai*.)

Z. Shao, Q. Zhai, and X. Guan are with Systems Engineering Institute, MOEKLINNS Lab, Xi'an Jiaotong University, Xi'an 710049, China. (e-mail: qzzhai@sei.xjtu.edu.cn).

form of iterative DC-PF model. Besides, the DC-PF models generally consider the branch losses as constants, and the values of the constant losses are from empirical estimation.

To improve the accuracy of DC-PF, the P-LPF models are proposed [14, 15]. The P-LPF models extend the basic idea of the classical DC-PF model and adopt the techniques of Taylor expansion. Compared with the DC-PF, the P-LPF models take into consideration the reactive power flow and voltage. Literature [16] summarizes different types of P-LPF models and proposes a general formulation for P-LPF models. In literature [17], the truncation error of the general P-LPF model is analyzed. In literature [18, 19], the P-LPF is further improved by expanding the variable space and some data-driven ideas are absorbed. The P-LPF models consider the branch losses by performing Taylor expansion on the AC-PF model around the base point and the branch losses are expressed as linear loss functions.

The D-LPF models are newly proposed, which break the limitations of variable space and the formulations of AC-PF model. The D-LPF models obtains the mappings between variables directly from historical data or measurement data. Literature [20] proposes a regression-based D-LPF model and uses the partial least-squares to address the data collinearity. Literature [21] proposes a iterative D-LPF approach to linearize the AC-PF equations considering the data noise. Literature [22] proposes a hybrid D-LPF model, in which the regression is applied to the errors of the P-LPF models. Literature [23] extends the data-driven idea to the convex approximation, and uses the ensemble learning method to find a quadratic mapping between the variables. In literature [24], the approximation error of the regression-based D-LPF model is bounded through the Rademacher complexity theory, in which the probability distribution is assumed. In literature [25], a theoretical explanation is given to illustrate why the performance of the D-LPF is much better than that of DC-PF and P-LPF. The D-LPF uses a double-ends formulation to consider the branch losses, i.e., the from-end and the to-end of the branch have different LPF models.

In the above literatures, the LPF models are generally evaluated by Monte Carlo (MC) simulation. However, the result of MC simulation is sensitive to the probability distribution of the samples, but the accurate probability distribution is difficult to obtain in engineering. Moreover, the MC simulation evaluates the LPF models in an expectation sense, and the accurate error bound of the LPF model is inaccessible.

Lots of LPF models are proposed and show excellent performance in their own settings, but to our best knowledge, no method is proposed to evaluate or rank them fairly. Therefore, it is meaningful to propose a fair and unified method to evaluate the performance of LPF models, and further, the method can tell whether a given branch is suitable for LPF or not, and what is the best LPF model for the given branch.

To this end, this paper proposes a general formulation to provide a unified and consistent evaluation system for the LPF models, under this system, the best LPF model can be defined and further obtained. The main contributions of this paper are summarized as follows:

1) The definition set of LPF models is proposed. The set can be adjusted according to applications, and the parameters of the set are easy to obtain in engineering.

2) The general formulation for evaluating the LPF models is proposed, and based on the general formulation, the best LPF model is obtained. A grid sampling method is proposed to give analytical solutions to the problems.

3) Under the proposed evaluation method, different LPF models are tested and ranked. The proposed evaluation method gives both the expected error and the maximum error of the LPF models. Besides, the factors that affect the performance of LPF models are analyzed.

The remainder of the paper is organized as follows: Section II provides the mathematical formulations, which includes the review of LPF models, the basic idea, and the methodology of the general formulation. Section III provides case studies and Section IV concludes the paper.

## II. MATHEMATICAL FORMULATION

In this paper, for all formulations, the phase angles are in radian, and other variables are in per-unit.

### A. The AC Branch Power Flow Model

The AC branch power flow models are provided as follows:
$$p_{ij} = g_l v_i^2 - v_i v_j g_l \cos\theta_{ij} - v_i v_j b_l \sin\theta_{ij}, \quad (1)$$
$$q_{ij} = -b_l v_i^2 + v_i v_j b_l \cos\theta_{ij} - v_i v_j g_l \sin\theta_{ij}.$$

The AC-PF model can be presented in a more compact from as follows (this treatment is possibly novel). For a given branch, it has:
$$g_l + jb_l = \frac{1}{r_l + jx_l} = \frac{1}{z_l e^{j\varphi_l}} = y_l e^{-j\varphi_l}. \quad (2)$$

Then, it can be obtained:
$$g_l = y_l \cos\varphi_l, \ b_l = -y_l \sin\varphi_l, \ \varphi_l = \arctan(-b_l/g_l). \quad (3)$$

By substituting (3) into the formulation (1), it has:
$$p_{ij} = g_l v_i^2 - v_i v_j y_l (\cos\varphi_l \cos\theta_{ij} - \sin\varphi_l \sin\theta_{ij}), \quad (4)$$
$$q_{ij} = -b_l v_i^2 - v_i v_j y_l (\sin\varphi_l \cos\theta_{ij} + \cos\varphi_l \sin\theta_{ij}).$$

By using the trigonometric formulas, the formulation (4) is transformed into the following formulation:
$$p_{ij} = g_l v_i^2 - v_i v_j y_l \cos(\theta_{ij} + \varphi_l), \quad (5)$$
$$q_{ij} = -b_l v_i^2 - v_i v_j y_l \sin(\theta_{ij} + \varphi_l).$$

### B. Formulations of Various Linear Power Flow Models

The formulations of three typical types of LPF models are provided in this subsection.

#### 1) The DC Power Flow Model

The DC-PF models only consider active power and can be summarized as the formulation (6).
$$p_{ij} = \gamma_{ij} \cdot \theta_{ij} + \alpha_{ij} \quad (6)$$

The values of $\gamma_{ij}$ and $\alpha_{ij}$ are generally case-dependent and empirical. In the classical DC-PF model, it has:
$$\gamma_{ij} = 1/x_l, \quad \alpha_{ij} = 0. \quad (7)$$

The detailed review of various DC-PF models is available in literature [7].

*2) The Physical-Model-Driven Linear Power Flow Model*

The P-LPF models are generally the transformation/simplification/Taylor expansion of AC-PF model. The typical P-LPF models are presented as follows:

$$p_{ij} = 0.5g_l(v_i^2 - v_j^2) - b_l\theta_{ij}, \qquad (8)$$
$$q_{ij} = -0.5b_l(v_i^2 - v_j^2) - g_l\theta_{ij}.$$

The formulation (8) is a lossless model. When the loss is considered, the P-LPF model should work in the hot-start mode and the base point is required.

More detailed summary of various P-LPF models is available in literature [16]. In [16], it is also announced that the P-LPF models can be generalized by expanding the variable space, for example, the $\ln v$ can be regarded as an independent variable to replace the $v^2$ in (8).

*3) The Data-Driven Linear Power Flow Model*

The D-LPF models mine the linear relationship between the variables from the data. The formulations of D-LPF models are flexible, and the typical D-LPF model of branch form can be expressed as follows:

$$p_{ij} = \alpha_{ij}^p + \sum_{n=1}^{N} \beta_n^{p_{ij}} v_n + \sum_{n=1}^{N} \gamma_n^{p_{ij}} \theta_n, \qquad (9)$$
$$q_{ij} = \alpha_{ij}^q + \sum_{n=1}^{N} \beta_n^{q_{ij}} v_n + \sum_{n=1}^{N} \gamma_n^{q_{ij}} \theta_n.$$

In the D-LPF model, the branch flows are related to voltages of all buses. The linear factors of D-LPF are obtained by linear regression (i.e., basically the least-squares method). The data of the D-LPF model is collected from historical data or real-time measurement data.

The D-LPF models are considered to perform better than the P-LPF and DC-PF models. In literature [25], it is explained that the good performance of D-LPF models comes from the fact that the D-LPF models are the best linear approximation on the *RS* (i.e., the regional set defined by the data), and the scenarios in the *RS* has a high probability of appearing in engineering. On the contrary, the D-LPF may perform worse than the P-LPF and DC-PF models on the *AS*. More detailed summary of the D-LPF models of branch form is available in [25].

*C. A General Formulation for Evaluating the Performance of Linear Power Flow Models*

Various types of LPF models are dedicated to finding a good linear approximation of AC-PF model. So which one is the best LPF model? In this subsection, we would like to answer this question and propose a general formulation (**GF**) to evaluating the performance of LPF models.

By observing the AC-PF model, it can be found that the branch power flow involves only four types of variables, which are $p_l/q_l$, $v_i$, $v_j$ and $\theta_{ij}$. Then, in practice, the four types of variables should be within a certain range, and combining these ranges, the definition range (DR) set of LPF models can be defined as the formulation (10).

In the formulation (10), the DR set considers both directions of the branch to ensure all scenarios in the DR set can satisfy the power flow limitations. The DR set is controllable by adjusting the parameters (e.g., the $\underline{V}_i, \overline{V}_i$), and in engineering, these parameters are easy to be obtained or estimated.

$$S_l = \left\{(p_l, q_l, v_i, v_j, \theta_{ij}) \middle| \begin{array}{l} p_{ij} = g_l v_i^2 - v_i v_j y_l \cos(\theta_{ij} + \varphi_l), \\ p_{ji} = g_l v_j^2 - v_i v_j y_l \cos(-\theta_{ij} + \varphi_l), \\ q_{ij} = -b_l v_i^2 - v_i v_j y_l \sin(\theta_{ij} + \varphi_l), \\ q_{ji} = -b_l v_j^2 - v_i v_j y_l \sin(-\theta_{ij} + \varphi_l), \\ \underline{V}_i \le v_i \le \overline{V}_i, \underline{V}_j \le v_j \le \overline{V}_j, \\ -\overline{F}_l^p \le p_{ij}, p_{ji} \le \overline{F}_l^p, \\ -\overline{F}_l^q \le q_{ij}, q_{ji} \le \overline{F}_l^q, \\ \underline{\theta}_{ij} \le \theta_{ij} \le \overline{\theta}_{ij}. \end{array}\right\} \quad (10)$$

The LPF models work within the set $S_l$ and are aimed to approximate the branch flows by linear functions of $v_i$, $v_j$ and $\theta_{ij}$. Therefore, the performance of an LPF model can be evaluated by the following formulations:

(**GP**):
$$\min_{\alpha_{ij}^p, \beta_i^{p_{ij}}, \beta_j^{p_{ij}}, \gamma_{ij}^p} \iiint \delta(p_{ij} - p_{ij}^{LPF}) d\theta_{ij} dv_j dv_i \qquad (11)$$

s.t. $p_{ij}^{LPF} = \alpha_{ij}^p + \beta_i^{p_{ij}}\chi(v_i) + \beta_j^{p_{ij}}\chi(v_j) + \gamma_{ij}^p\phi(\theta_{ij})$ (12)

$$p_{ij}, v_i, v_j, \theta_{ij} \in S_l \qquad (13)$$

(**GQ**):
$$\min_{\alpha_{ij}^q, \beta_i^{q_{ij}}, \beta_j^{q_{ij}}, \gamma_{ij}^q} \iiint \delta(q_{ij} - q_{ij}^{LPF}) d\theta_{ij} dv_j dv_i \qquad (14)$$

s.t. $q_{ij}^{LPF} = \alpha_{ij}^q + \beta_i^{q_{ij}}\chi(v_i) + \beta_j^{q_{ij}}\chi(v_j) + \gamma_{ij}^q\phi(\theta_{ij})$ (15)

$$q_{ij}, v_i, v_j, \theta_{ij} \in S_l \qquad (16)$$

Where, the constraints (12) and (15) represent the general form of LPF models, for example, the P-LPF model in (8) can be generalized as:

$$\chi(v) = v^2, \ \phi(\theta) = \theta, \ \alpha_{ij}^p = 0, \qquad (17)$$
$$\beta_i^{p_{ij}} = -\beta_j^{p_{ij}} = 0.5g_l, \ \gamma_{ij}^p = -b_l.$$

The error metric in (11) and (14) is determined according to applications. For example, it can be the absolute error, the least-squares or the minimization of the maximum error.

The **GF** have three functions:
- The **GF** is able to provide the best LPF (B-LPF) model on the DR set.
- The **GF** is able to judge whether the branch is suitable for LPF models by analyzing the error of the B-LPF.
- When the LPF model is given, the **GF** is able to evaluate the performance of the LPF on the DR set.

It should be noted that the **GF** has non-convex constraints and integral objectives, which cannot be solved directly. Therefore, an analytical method is proposed to solve the problems of **GF**.

*D. Problem for Obtaining the Best Linear Power Flow Model*

In this section, the **GF** will be instantiated to construct an optimization problem for obtaining the best LPF on the DR set. The least-squares is selected as the error metric, and the settings in formulation (18) are used.

$$\delta(\cdot) = (\cdot)^2, \ \chi(v) = v^2, \ \phi(\theta) = \theta \qquad (18)$$

With the settings in (18), the problem for obtaining the B-LPF model can be described as follows:

**(GP-I)**:

$$\min_{\alpha_{ij}^p, \beta_i^{p_{ij}}, \beta_j^{p_{ij}}, \gamma_{ij}^p} \left( p_{ij}^{(k)} - p_{ij}^{LPF,(k)} \right)^2 \quad (19)$$

s.t. $p_{ij}^{LPF,(k)} = \alpha_{ij}^p + \beta_i^{p_{ij}}(v_i^{(k)})^2 + \beta_j^{p_{ij}}(v_j^{(k)})^2 + \gamma_{ij}^p \theta_{ij}^{(k)}, \forall k$ (20)

$$p_{ij}^{(k)}, v_i^{(k)}, v_i^{(k)}, \theta_{ij}^{(k)} \in S_l \quad (21)$$

**(GQ-I)**:

$$\min_{\alpha_{ij}^q, \beta_i^{q_{ij}}, \beta_j^{q_{ij}}, \gamma_{ij}^q} \left( q_{ij}^{(k)} - q_{ij}^{LPF,(k)} \right)^2 \quad (22)$$

s.t. $q_{ij}^{LPF,(k)} = \alpha_{ij}^q + \beta_i^{q_{ij}}(v_i^{(k)})^2 + \beta_j^{q_{ij}}(v_j^{(k)})^2 + \gamma_{ij}^q \theta_{ij}^{(k)}, \forall k$ (23)

$$q_{ij}^{(k)}, v_i^{(k)}, v_i^{(k)}, \theta_{ij}^{(k)} \in S_l \quad (24)$$

In the above formulations, the $S_l$ is a continuous set, which contains countless scenarios, and this makes the problems **GP-I** and **GQ-I** have infinite constraints. To make the problems solvable, representative scenarios in the $S_l$ are selected by using a grid sampling method.

Let, $F_{ij}^P(v_i, v_j, \theta_{ij}) = g_l v_i^2 - v_i v_j y_l \cos(\theta_{ij} + \varphi_l),$ (25)

$F_{ij}^Q(v_i, v_j, \theta_{ij}) = -b_l v_i^2 - v_i v_j y_l \sin(\theta_{ij} + \varphi_l).$ (26)

Then, the grid sampling method (**GSM**) can be described as follows:

---

**Grid Sampling Method**

**Step 1**: (*Initialization*) Set the parameters $B$, $D$ and $M$, then generate $v_i^{(b)}, v_j^{(d)}, \theta_{ij}^{(m)}$ by the following formulations:

$v_i^{(b)} = \underline{V}_i + [(\overline{V}_i - \underline{V}_i)(b-1)]/(B-1), \forall b=1,2,\cdots,B$ (27)

$v_j^{(d)} = \underline{V}_j + [(\overline{V}_j - \underline{V}_j)(d-1)]/(D-1), \forall d=1,2,\cdots,D$ (28)

$\theta_{ij}^{(m)} = \underline{\theta}_{ij}^{Re} + [(\overline{\theta}_{ij}^{Re} - \underline{\theta}_{ij}^{Re})(m-1)]/(M-1), \forall m=1,2,\cdots,M$ (29)

**Step 2**: (*Sampling*) Set $k=0$ and initialize $S_l = \emptyset$. Then do the following procedures:

For $b$ from 1 to $B$,
  For $d$ from 1 to $D$,
    For $m$ from 1 to $M$,
      Calculate the branch flows by:
      $p_{ij}^* = F_{ij}^P(v_i^{(b)}, v_j^{(d)}, \theta_{ij}^{(m)}), p_{ji}^* = F_{ij}^P(v_j^{(d)}, v_i^{(b)}, -\theta_{ij}^{(m)}),$
      $q_{ij}^* = F_{ij}^Q(v_i^{(b)}, v_j^{(d)}, \theta_{ij}^{(m)}), q_{ji}^* = F_{ij}^Q(v_j^{(d)}, v_i^{(b)}, -\theta_{ij}^{(m)}).$
      If $-\overline{F}_l^p \le p_{ij}^*, p_{ji}^* \le \overline{F}_l^p$ & $-\overline{F}_l^q \le q_{ij}^*, q_{ji}^* \le \overline{F}_l^q$
        save $(p_{ij}^*, p_{ji}^*, q_{ij}^*, q_{ji}^*, v_i^{(b)}, v_j^{(d)}, \theta_{ij}^{(m)})$ to $S_l$,
        $k=k+1$;
      End if.
    End for.
  End for.
End for.

---

It is worth noting that the range of $\theta_{ij}$ in the formulation (29) is $[\underline{\theta}_{ij}^{Re}, \overline{\theta}_{ij}^{Re}]$, which represents the real range of $\theta_{ij}$, and the real range can be defined as follows:

$$\overline{\theta}_{ij}^{Re} = \max \theta_{ij} \quad \text{and} \quad \underline{\theta}_{ij}^{Re} = \min \theta_{ij} \quad (30)$$
$$\text{s.t.} \quad \theta_{ij} \in S_l. \quad\quad \text{s.t.} \quad \theta_{ij} \in S_l.$$

In this paper, three types of the $\theta_{ij}$ ranges are involved, which are:

1) The roughly estimated range of $\theta_{ij}$, i.e., $\theta_{ij} \in [\underline{\theta}_{ij}, \overline{\theta}_{ij}]$.

2) The implicit range of $\theta_{ij}$, i.e., $\theta_{ij} \in [\underline{\theta}_{ij}^{Im}, \overline{\theta}_{ij}^{Im}]$, which comes from the enforcement of active power limitations.

3) The real range of $\theta_{ij}$, i.e., $\theta_{ij} \in [\underline{\theta}_{ij}^{Re}, \overline{\theta}_{ij}^{Re}]$.

A smaller range of $\theta_{ij}$ can help to improve the sampling accuracy of the **GSM**. For the active branch flow from bus $i$ to bus $j$, the $[\underline{\theta}_{ij}^{Re}, \overline{\theta}_{ij}^{Re}]$ can be obtained by the following analytical formulas:

$$\underline{\theta}_{ij}^{Re} = \max\{\underline{\theta}_{ij}^{Im}, \underline{\theta}_{ij}\}, \quad \overline{\theta}_{ij}^{Re} = \min\{\overline{\theta}_{ij}^{Im}, \overline{\theta}_{ij}\}. \quad (31)$$

$$\overline{\theta}_{ij}^{Im} / \underline{\theta}_{ij}^{Im} = \max/\min \left[ \arccos\left( \frac{g_l v_i^2 - p_{ij}}{v_i v_j y_l} \right) - \varphi_l \right] \quad (32)$$

s.t. $p_{ij} \in \{-\overline{F}_{ij}^p, \overline{F}_{ij}^p\}, v_i \in \{\underline{V}_i, \overline{V}_i\}, v_j \in \{\underline{V}_j, \overline{V}_j\}.$

The formulations of (32) comes from the analysis of the AC-PF partial derivatives. With the help of the compact AC-PF model, the partial derivatives can be easily analyzed, and the results are shown as follows:

Let $F(v_i, v_j, \theta_{ij}) = g_l v_i^2 - v_i v_j y_l \cos(\theta_{ij} + \varphi_l),$ (33)

Then it has,

$$\frac{\partial \theta_{ij}}{\partial v_j} = -\frac{F_{v_j}}{F_\theta} = -\frac{-v_i y_l \cos(\theta_{ij} + \varphi_l)}{v_i v_j y_l \sin(\theta_{ij} + \varphi_l)} = \frac{\cot(\theta_{ij} + \varphi_l)}{v_j} \quad (34)$$

The (34) indicates that the $v_j$ is monotonic to $\theta_{ij}$, and the optimal value of $\theta_{ij}$ is obtained at $v_j \in \{\underline{V}_j, \overline{V}_j\}$.

$$\frac{\partial \theta_{ij}}{\partial v_i} = -\frac{F_{v_i}}{F_\theta} = -\frac{2\cos\varphi_{ij}}{v_j \sin(\theta_{ij} + \varphi_{ij})} + \frac{\cos(\theta_{ij} + \varphi_{ij})}{v_i \sin(\theta_{ij} + \varphi_{ij})} \quad (35)$$

For common branches, it takes the assumptions of:

$$\theta_{ij} \approx 0, \varphi_{ij} \in [\pi/3, \pi/2], v_i, v_j \in [0.9, 1.1]. \quad (36)$$

Then the extreme point from the (35) is:

$$v_i^* = \frac{\cos(\theta_{ij} + \varphi_{ij})}{2\cos\varphi_{ij}} v_j \approx 0.5 v_j \notin [0.9, 1.1] \quad (37)$$

The (37) indicates that the extreme point of $v_i$ is out of the range of $v_i$, and the optimal solution of $\theta_{ij}$ is only obtained at the boundary of $v_i \in \{\underline{V}_i, \overline{V}_i\}$.

Q.E.D

Two points are worth attentions:

1) The algorithm in the formulation (32) should be applied to both $p_{ij}$ and $p_{ji}$, and the intersection of the two implicit ranges from $p_{ij}$ and $p_{ji}$ is the final real range of $\theta_{ij}$.

2) The range of $\theta_{ij}$ is sensitive to the active power but not sensitive to the reactive power. Therefore, the implicit range of $\theta_{ij}$ only comes from the limitations of active branch flow.

With the **GSM**, the set $S_l$ becomes $S_l(K)$, which is a set containing $K$ scenarios. Then, the problems **GP-I** and **GQ-I** are solvable, and the analytical solution of **GP-I** and **GQ-I** are given as follows:

Let $\boldsymbol{x}^{(k)} = (1, (v_i^{(k)})^2, (v_j^{(k)})^2, \theta_{ij}^{(k)})^T, \boldsymbol{y}^{(k)} = (p_{ij}^{(k)}, q_{ij}^{(k)})^T,$ (38)

and $\boldsymbol{X} = (\boldsymbol{x}^{(1)}, \boldsymbol{x}^{(2)}, \cdots, \boldsymbol{x}^{(K)}), \boldsymbol{Y} = (\boldsymbol{y}^{(1)}, \boldsymbol{y}^{(2)}, \cdots, \boldsymbol{y}^{(K)}),$ (39)

$$\boldsymbol{A} = \begin{pmatrix} \alpha_{ij}^p & \beta_i^{p_{ij}} & \beta_j^{p_{ij}} & \gamma_{ij}^p \\ \alpha_{ij}^q & \beta_i^{q_{ij}} & \beta_j^{q_{ij}} & \gamma_{ij}^q \end{pmatrix}^T. \quad (40)$$

Then, the solution is:

$$\boldsymbol{A} = (\boldsymbol{XX}^T)^{-1} \boldsymbol{XY}^T \quad (41)$$

*E. Problem for Evaluating the Performance of Linear Power Flow Models*

For a given LPF model, the performance (i.e., the error bound) of the LPF model can be obtained by the following problems:

**(GP-II)**

$$\min_{p_{ij}^{(k)}, v_i^{(k)}, v_j^{(k)}, \theta_{ij}^{(k)}} E_{ij}^p \quad (42)$$

s.t. $\quad -E_{ij}^p \leq p_{ij}^{(k)} - p_{ij}^{LPF,(k)} \leq E_{ij}^p, \quad \forall k \quad (43)$

$$p_{ij}^{LPF,(k)} = \alpha_{ij}^{p*} + \beta_i^{p_{ij}*}\chi(v_i^{(k)}) + \beta_j^{p_{ij}*}\chi(v_j^{(k)}) + \gamma_{ij}^{p*}\phi(\theta_{ij}^{(k)}), \forall k \quad (44)$$

$$p_{ij}^{(k)}, v_i^{(k)}, v_j^{(k)}, \theta_{ij}^{(k)} \in S_l(K) \quad (45)$$

**(GQ-II)**

$$\min_{q_{ij}^{(k)}, v_i^{(k)}, v_j^{(k)}, \theta_{ij}^{(k)}} E_{ij}^q \quad (46)$$

s.t. $\quad -E_{ij}^q \leq q_{ij}^{(k)} - q_{ij}^{LPF,(k)} \leq E_{ij}^q, \quad \forall k \quad (47)$

$$q_{ij}^{LPF,(k)} = \alpha_{ij}^{q*} + \beta_i^{q_{ij}*}\chi(v_i^{(k)}) + \beta_j^{q_{ij}*}\chi(v_j^{(k)}) + \gamma_{ij}^{q*}\phi(\theta_{ij}^{(k)}), \forall k \quad (48)$$

$$q_{ij}^{(k)}, v_i^{(k)}, v_j^{(k)}, \theta_{ij}^{(k)} \in S_l(K) \quad (49)$$

The error bounds of the given LPF model are obtained by the problems of **GP-II** and **GQ-II**. Actually, the obtained error bound is an estimation of the true error bound. Since the sampling density of the DR set is controllable, the estimated error bound is already very close to the true bound.

The **GP-II** and **GQ-II** can be solved analytically through the enumeration of the scenarios in the $S_l(K)$. By using the enumeration, not only the maximum error but also the average error of the LPF model can be obtained. The DR set is fully covered by the sampled scenarios, therefore, the error results from the above problems can be a fair and consistent index for evaluating the LPF models.

The flowchart of the proposed method is shown in Fig. 1.

*F. Discussions of the Proposed Method*

*1) The computational burden and accuracy*

The solution of the problems, including the GP-I, GQ-I, GP-II, and GQ-II, are analytical. Therefore, the solution speed of the problems is very fast. For a single branch, the entire process (i.e., including the sampling and the calculation of B-LPF) under the settings of $B = D = M = 100$ (i.e., 1 million samples in the DR set) requires only about 1.6 seconds. Under such solution speed, even in the face of large-scale systems who have thousands of branches, the solution time of the proposed method is still competent.

The accuracy of the proposed method is controllable. For example, when the voltage range is set to $v_i, v_j \in [0.9, 1.1]$, and $B = D = 100$, then, the accuracy of the voltage is 0.002 p.u..

*2) The losses of the branch*

The proposed B-LPF model adopts the double-end modeling method, that is, the $P_{ij}$ and $p_{ji}$ have different model factors, so that the branch losses can be approximated by the following linear functions:

$$p_l^{loss} = p_{ij} + p_{ji}$$
$$= (\alpha_{ij}^p + \alpha_{ji}^p) + (\beta_i^{P_{ij}} + \beta_i^{P_{ji}})v_i$$
$$+ (\beta_j^{P_{ij}} + \beta_j^{P_{ji}})v_j + (\gamma_{ij}^p - \gamma_{ji}^p)\theta_{ij} \quad (50)$$

$$q_l^{loss} = q_{ij} + q_{ji}$$
$$= (\alpha_{ij}^q + \alpha_{ji}^q) + (\beta_i^{q_{ij}} + \beta_i^{q_{ji}})v_i$$
$$+ (\beta_j^{q_{ij}} + \beta_j^{q_{ji}})v_j + (\gamma_{ij}^q - \gamma_{ji}^q)\theta_{ij} \quad (51)$$

*3) Visualization of the AC-PF model.*

The AC-PF model is usually regarded as a hypersurface that is difficult to visualize. With the compact form of AC-PF model, a visualization technique is given here. The technique shows the AC-PF model on the $P_{ij}$ - $\theta_{ij}$ plane. The active branch flow of an example branch is show in the Fig. 2, in which the $AS^P$ and $RS^P$ represent the $AS$ and $RS$ that only related to the active power. The example branch has $g = 14.14$, $b = -45.62$, $\varphi = 1.27$, $\bar{F}^p = 1.08$.

From the Fig. 2, it can be found that the AC-PF model is a series of cosine-like functions that are stretched and translated by the voltage magnitudes. The visualization of AC-PF model is helpful in the analysis of the characteristics of the LPF model.

*4) The factors that impact the approximation accuracy of LPF models*

Generally, a multivariate function is close to a linear function when the Hessian matrix is close to zero. Based on this proposal, the Hessian matrix of AC-PF model is shown in the Table I.

With the compact form of the AC-PF model, the Hessian matrix is easier to be analyzed. From the results in the Table II, it can be found that the values of $y_l$, $\theta_{ij}$ (i.e., essentially reflects the value of $F_l^p$) and $\varphi_l$ (i.e., essentially reflects the value of $r_l/x_l$) are the key factors that affect the approximation accuracy of the AC-PF model.

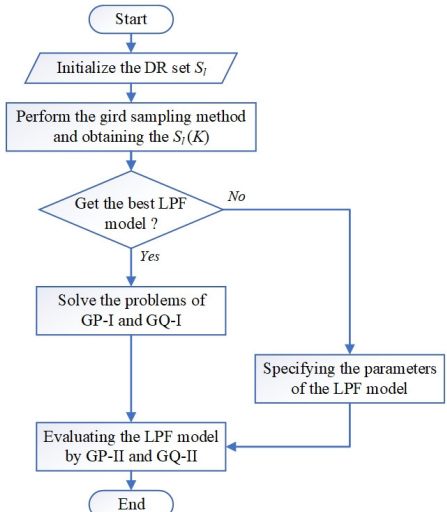

Fig.1 The flowchart of the proposed method.

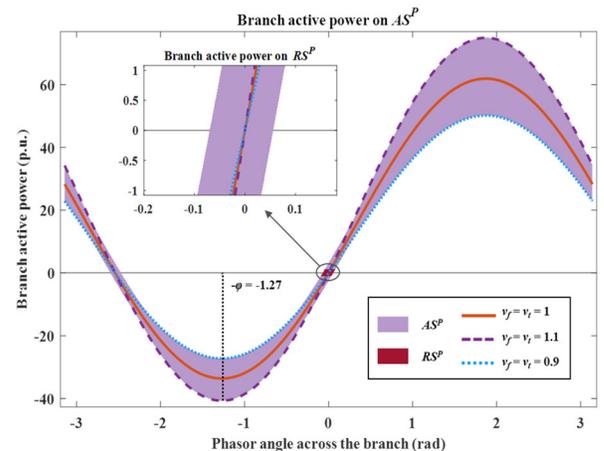

Fig.2 The example of the visualization of the AC-PF model.

TABLE I
THE HESSIAN MATRICES OF THE AC-PF MODEL

| $H^P$ | $p_{ij}$ | | |
|---|---|---|---|
| | versus $\theta_{ij}$ | versus $v_i$ | versus $v_j$ |
| versus $\theta_{ij}$ | $v_i v_j y_l \cos(\theta_{ij}+\varphi_l)$ | $v_j y_l \sin(\theta_{ij}+\varphi_l)$ | $v_i y_l \sin(\theta_{ij}+\varphi_l)$ |
| versus $v_i$ | $v_j y_l \sin(\theta_{ij}+\varphi_l)$ | $2g_l$ | $-y_l \cos(\theta_{ij}+\varphi_l)$ |
| versus $v_j$ | $v_i y_l \sin(\theta_{ij}+\varphi_l)$ | $-y_l \cos(\theta_{ij}+\varphi_l)$ | 0 |
| $H^Q$ | $q_{ij}$ | | |
| | versus $\theta_{ij}$ | versus $v_i$ | versus $v_j$ |
| versus $\theta_{ij}$ | $v_i v_j y_l \sin(\theta_{ij}+\varphi_l)$ | $-v_j y_l \cos(\theta_{ij}+\varphi_l)$ | $-v_i y_l \cos(\theta_{ij}+\varphi_l)$ |
| versus $v_i$ | $-v_j y_l \cos(\theta_{ij}+\varphi_l)$ | $-2b_l$ | $-y_l \sin(\theta_{ij}+\varphi_l)$ |
| versus $v_j$ | $-v_i y_l \cos(\theta_{ij}+\varphi_l)$ | $-y_l \sin(\theta_{ij}+\varphi_l)$ | 0 |

## III. CASE STUDY

### A. Basic Information

Several standard systems from the MATPOWER 7.1 [26] are tested. The systems are IEEE-24-bus system, 30-bus system, 118-bus system, ACTIVSg-200-bus system, ACTIVSg-500-bus system, ACTIVSg-2000-bus system.

LPF models are tested and compared, which are the proposed LPF model (denoted as B-LPF), the P-LPF model in (8), the D-LPF model with the following formulation of (52), and the classical DC-PF model.

$$p_{ij}^{D-LPF} = \alpha_{ij}^p + \beta_i^{p_{ij}} v_i^2 + \beta_j^{p_{ij}} v_j^2 + \gamma_{ij}^p \theta_{ij}$$
$$q_{ij}^{D-LPF} = \alpha_{ij}^q + \beta_i^{q_{ij}} v_i^2 + \beta_j^{q_{ij}} v_j^2 + \gamma_{ij}^q \theta_{ij}$$
(52)

The data used in the D-LPF model is provided by the simulation approach in literature [25]. The load in the simulation process is revised according to a real annual curve of 8760h.

### B. The Comparison Results and Overall Performance of the LPF Models

The LPF models are compared by using the proposed evaluation method. The set $S_l$ in the comparison is configured as the Table II.

The parameters in the grid sampling method are set to $B = D = M = 100$, which means that the maximum number of samples is one million and the voltage accuracy is 0.002 p.u.. It is worth attention that the phase angle accuracy is branch-dependent.

The comparison results are show in the Table III. Since the error of the branch with larger capacity are often larger, to make the results fair, the error is normalized by the branch capacity, which can be described as the following formulation:

$$e_{ij} = \frac{\left|p_{ij} - p_{ij}^{LPF}\right|}{\bar{F}_{ij}^p} \times 100\%. \quad (53)$$

From the results in the Table III, three points can be founded as follows:

*1) The Rank of the Performance of the LPF Models*

From the results in the Table III, it can be found that the performance rank of the LPF models (i.e., from the accurate one to the inaccurate one) is B-LPF > P-LPF > DC-PF > D-LPF. This result is expected. The performance of B-LPF is the best, which confirms the theory of this paper.

Besides, it is found that the performance of P-LPF is closer to that of the B-LPF, which shows that the P-LPF model is also a qualified model. The DC-PF, as an early LPF model, has a relatively good performance in the average sense (i.e., the Avg. Error), but the DC-PF has a relatively large error in the maximum sense (i.e., Max. Error), and the maximum error of DC-PF among all test systems reaches 506.5%, which is totally unacceptable in engineering. Over all, the DC-PF performance under the proposed evaluation method is consistent with the empirical knowledge of the DC-PF model.

Particularly, the performance of the D-LPF model seems to be abnormal, because it is announced in literatures that the D-LPF model is much better than the P-LPF and DC-PF. In fact, this result is reasonable, and it just confirms the theory that we put forward in literature [25], in which the theory is that the D-LPF model is actually the near-optimal LPF model on the *RS*, and the approximation error of the D-LPF is likely to be greater when performing the D-LPF on the *AS*. The DR set used in the comparison represents a quite wide range, which can be regarded as the *AS*, therefore, it is theoretically reasonable for the D-LPF to perform poorly on this DR set.

TABLE II
CONFIGURATIONS OF THE SET

| $[\underline{V}_i, \bar{V}_i]$ | $[\underline{V}_j, \bar{V}_j]$ | $[\underline{\theta}_{ij}, \bar{\theta}_{ij}]$ | $\bar{F}_{ij}^p$ | $\bar{F}_{ij}^q$ |
|---|---|---|---|---|
| [0.9, 1.1] | [0.9, 1.1] | [-π/3, π/3] | Default | Equal to $\bar{F}_{ij}^p$ |

TABLE III
COMPARISON RESULTS OF THE LPF MODELS

| Systems | Active Power | | | | | | | | Reactive Power | | | | | | |
|---|---|---|---|---|---|---|---|---|---|---|---|---|---|---|---|
| | Max. Error (%) | | | | Avg. Error (%) | | | | Max. Error (%) | | | Avg. Error (%) | | | |
| | B-LPF | P-LPF | D-LPF | DC | B-LPF | P-LPF | D-LPF | DC | B-LPF | P-LPF | D-LPF | B-LPF | P-LPF | D-LPF | |
| 24-bus | 28.4 | 36.5 | 56.7 | 43.1 | 3.6 | 3.8 | 6.1 | 7.0 | 25.7 | 38.2 | 100.4 | 3.3 | 5.2 | 11.2 | |
| 30-bus | 23.8 | 28.8 | 124.1 | 153.6 | 3.9 | 4.1 | 9.9 | 18.7 | 21.0 | 25.7 | 123.7 | 2.2 | 2.9 | 7.5 | |
| 118-bus | 37.6 | 106.1 | 276.5 | 111.4 | 3.6 | 4.6 | 8.9 | 10.8 | 41.0 | 110.9 | 411.7 | 3.0 | 6.2 | 13.8 | |
| 200-bus | 25.6 | 31.7 | 506.5 | 42.4 | 3.9 | 4.1 | 19.1 | 6.8 | 20.3 | 29.6 | 281.9 | 2.0 | 3.3 | 13.3 | |
| 500-bus | 24.5 | 30.3 | 302.8 | 46.3 | 4.1 | 4.4 | 26.5 | 6.3 | 19.5 | 28.1 | 175.5 | 1.6 | 2.6 | 14.1 | |
| 2000-bus | 48.4 | 99.6 | 450.6 | 149.1 | 4.0 | 4.2 | 16.6 | 8.1 | 72.7 | 110.4 | 1375.7 | 1.8 | 2.9 | 7.7 | |
| Total Avg. | 31.4 | 55.5 | 286.2 | 91.0 | 3.9 | 4.2 | 14.5 | 9.6 | 33.4 | 57.2 | 411.5 | 2.3 | 3.8 | 11.3 | |

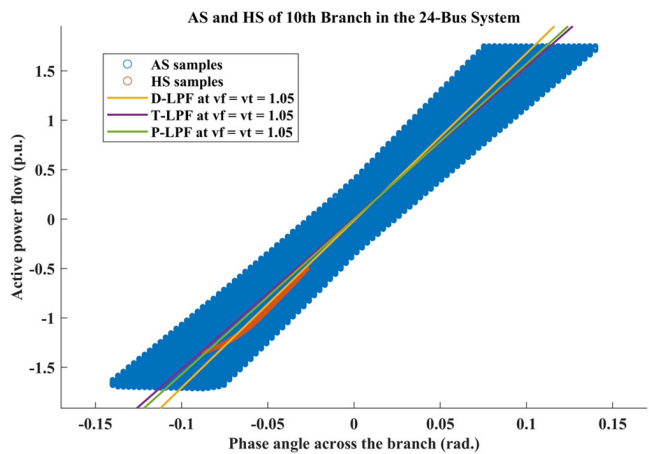

Fig.3 Example of the *HS* and the *AS*.

TABLE IV
THE LPF FACTORS OF THE 10$^{TH}$ BRANCH IN 24-BUS SYSTEM

| Models | \multicolumn{4}{c}{Factors of the LPF models} |

| Models | $\alpha_{ij}^{p}$ | $\beta_{i}^{p_{ij}}$ | $\beta_{j}^{p_{ij}}$ | $\gamma_{ij}^{p}$ |
|---|---|---|---|---|
| B-LPF | 0.0193 | 1.8033 | -1.8104 | 15.2805 |
| P-LPF | 0 | 1.8036 | -1.8036 | 15.7002 |
| D-LPF | 1.3822 | 1.4064 | -2.6735 | 16.9041 |
| DC-PF | 0 | / | / | 16.5289 |
| \multicolumn{5}{c}{Parameters of the branch} | | | | |
| $y$ | $g$ | $b$ | $\varphi$ | $\bar{F}^{P} / \bar{F}^{Q}$ |
| 16.11 | 3.61 | -15.70 | 1.345 | 1.75 |

*2) The D-LPF is a Special Case of the B-LPF*

To further illustrate the result of D-LPF, it takes the 10$^{th}$ branch of 24-bus system as an example to give a visualization of *AS* (i.e., the $S_l$) and *HS*. The 10$^{th}$ branch is selected since it has the maximum Max. Error in the 24-bus system. The *AS* and *HS* of the active branch flow of the example branch are shown in the Fig. 3.

In the Fig. 3, it can be seen that the *HS* is a small subset of the *AS*, and obviously, the D-LPF performs better than the B-LPF and P-LPF on the *HS* but worse on the *AS*. The subset *HS* is an insufficient representation of the *AS*, which may lead to factors' difference between the B-LPF and the D-LPF. The Table IV shows the factors of the four LPF models as well as the parameters of the 10$^{th}$ branch. From the Table IV, it can be seen that the values of the factors of B-LPF and P-LPF are close, but the D-LPF are different from other LPF models.

In fact, the D-LPF model is a special case of the B-LPF model, this is because by reasonably adjusting the parameters of $S_l$, the *HS* can be represented by the $S_l$. Then the B-LPF model under this *HS* is just the D-LPF model. In this way, the D-LPF can be regarded as a special case of the B-LPF.

*3) The Significant Difference Between the Avg. Error and Max. Error*

It can be found in the Table III that the difference between the Avg. Error and the Max. Error is very large (e.g., observing the last row of Table III, the Avg. Error of B-LPF is 3.9% and the Max. Error of B-LPF is 31.4%). This phenomenon actually shows a fact that most of the branches in real power systems can be well approximated by an LPF model, and a small part of the branches are naturally impossible to be approximated.

In the past, the question of whether a branch is applicable for an LPF model is generally answered empirical.

In this paper, whether a branch is suitable for a LPF model can be judged through the performance of the B-LPF. In theory, the B-LPF is the best LPF model on the given $S_l$, and if the approximation error of B-LPF exceeds the acceptable criterion, it can be determined that this branch is not suitable for linear approximation. This evaluation is simple and easy to implement in engineering.

*C. The Error Distribution of the LPF Models*

The significant difference between the average error and the maximum error reminds us to pay attention to the error distributions. In this subsection, the error distributions are shown by using the 24-bus, 118-bus and 500-bus systems.

The error distributions of the active and reactive branch flows are shown in the Fig. 4 and Fig. 5 respectively. From the Fig. 4 and Fig. 5, it can be found that:

1) The error distributions of active and reactive power of B-LPF and P-LPF models are basically the same. About 1/3 of the errors are less than 1%.

2) The D-LPF and DC-PF models have relatively large errors, and about 80% errors are within the large error range (i.e., the error > 5%).

3) It is worth noting that the Max. Errors in the Table III appear rarely in the distributions. For example, the Max. Error of P-LPF model on 118-bus system is 106.1%. But in the Fig. 4, the errors of P-LPF that exceeding 25% are rare and occupy

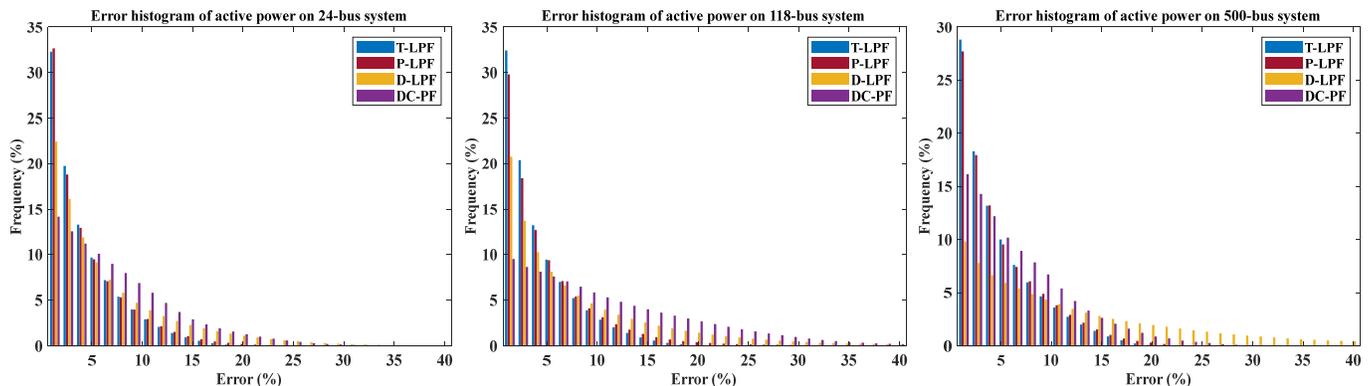

Fig. 4 Error distribution of active power on 24-bus, 118-bus, 500-bus systems.

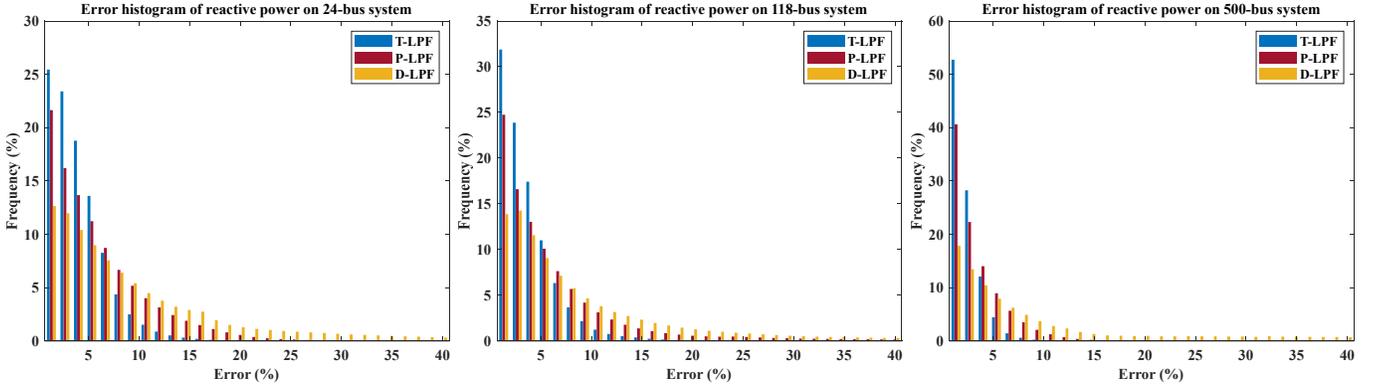

Fig. 5 Error distribution of reactive power on 24-bus, 118-bus, 500-bus systems.

only a small part. This phenomenon confirms the robustness of the proposed LPF model.

4) From the error distributions of B-LPF and P-LPF, it can be seen that the B-LPF is slightly better than the P-LPF in the active power, and the B-LPF is significantly better than the P-LPF as for the reactive power.

From the above results, it can be concluded that the B-LPF performs the best among the four types of LPF models, and the low frequency of large errors verifies the robustness and effectiveness of the B-LPF model.

### D. The Approximation Accuracy of Branch Losses

The P-LPF model and DC-PF model considers the branch losses in a hot-start mode, which needs an empirical base point. On the contrary, the B-LPF model are able to approximate the losses in a cold-start mode.

In this subsection, the approximation accuracy of branch losses of the B-LPF model is investigated. The errors of losses are normalized by the following formulation:

$$e_l^{loss} = \frac{\left| p_l^{loss} - p_l^{LPF,loss} \right|}{\max\{p_l^{loss,(k)}, \forall k \in S_l\}} \times 100\% \quad (54)$$

The approximation results of the branch losses are show in the Table V. From the results in the Table V, it can be found that the approximation errors of active losses and reactive losses are exactly the same. This phenomenon actually shows that the B-LPF model keeps the physical laws when approximating the losses. In the AC-PF model, the active losses and reactive losses always keep the following linear relationship:

$$\dot{s}_l^{loss} = p_l^{loss} + jq_l^{loss} = (\dot{y}_l \dot{v}_l)^* \dot{v}_l = g_l v_l^2 - jb_l v_l^2$$
$$\Downarrow \quad (55)$$
$$p_l^{loss} / q_l^{loss} = -g_l / b_l = \text{Constant}$$

Where, $\dot{y}_l$ is the admittance phasor of branch $l$, $\dot{v}_l$ is the voltage phasor difference across the branch $l$.

In different test systems, the approximate errors of the branch losses are very close, the Max. Err of all systems are about 75.1%, and the Avg. Err are about 16.6%. This result shows that B-LPF is effective and robust in the approximation of branch losses.

### E. The Factors that Impact the Approximation Accuracy

The method in this paper provides a large amount of error data, and the error data is not disturbed by the randomness of sampling. Therefore, we would like to analyze the key factors by calculating the correlation coefficients (denoted as $r_c$) between the absolute errors of the B-LPF and the various branch parameters. The results of the correlation coefficients are shown in the Table VI.

TABLE V
ERRORS OF BRANCH LOSSES OF THE B-LPF MODEL

| Systems | Active Power | | Reactive Power | |
|---|---|---|---|---|
| | Max. Err (%) | Avg. Err (%) | Max. Err (%) | Avg. Err (%) |
| 24-bus | 73.5 | 17.3 | 73.5 | 17.3 |
| 30-bus | 72.6 | 15.9 | 72.6 | 15.9 |
| 118-bus | 73.9 | 16.1 | 73.9 | 16.1 |
| 200-bus | 74.0 | 16.5 | 74.0 | 16.5 |
| 500-bus | 82.9 | 17.7 | 82.9 | 17.7 |
| 2000-bus | 74.0 | 16.1 | 74.0 | 16.1 |
| Total Avg. | 75.1 | 16.6 | 75.1 | 16.6 |

Generally, it is considered to be irrelevant when the $r_c \leq 0.3$; it is considered to be a weak correlation when the $0.3 < r_c \leq 0.8$, and it is considered to be a strong correlation when the $0.8 < r_c \leq 1$.

In Table VI, it can be found that the power flow limitation $\bar{F}$ is the most crucial factor, i.e., branches with large capacity generally have larger absolute errors, in fact, the value of $\bar{F}$ is directly related to the range of $\theta_{ij}$, and from the Hessian matrix in the Table I, it is evident that the $\theta_{ij}$ is the key variable to affect the values of the second-order partial derivatives. The second factor is the impedance angle $\varphi$, which represents the value of $r/x$, and empirically, the larger value of $r/x$ leads to larger approximation errors. Furthermore, the values of $y$ and $b$ show a weak correlation on the approximation errors. On the whole, the above results validate the factors that are analyzed in the Section II-F.

TABLE VI
CORRELATION COEFFICIENTS BETWEEN ABSOLUTE ERRORS
AND BRANCH PARAMETERS

| Correlations | $y$ | $g$ | $b$ | $\varphi$ | $\bar{F}$ |
|---|---|---|---|---|---|
| P-MaxErr | 0.29 | 0.16 | 0.29 | 0.45 | 0.96 |
| Q-MaxErr | 0.21 | 0.25 | 0.21 | 0.33 | 0.77 |
| P-AvgErr | 0.37 | 0.22 | 0.38 | 0.45 | 0.95 |
| Q-AvgErr | 0.37 | 0.22 | 0.38 | 0.45 | 0.95 |

## IV. Conclusion

LPF models are very important in power system analysis. Although many LPF models are proposed, some key questions related to the LPF models are still remained, i.e., what is the performance bound of the LPF model, which LPF model is the most suitable one for a given branch, and which branch cannot be linear approximated. To answer these questions, this paper proposes a general formulation to give a simple, fair and uniform evaluation standard for the LPF models. Based on the general formulation, the problems for obtaining the best LPF model and evaluating the given LPF model are proposed. The analytical solutions for the problems are also provided. The method in this paper is simple and easy to implement, which is valuable in engineering applications.